\documentclass{jpsj2}
\usepackage{graphicx}
\usepackage{amsmath}

\def\runtitle{%
Josephson effect in d-wave superconductor junctions in a
lattice model}  
\def\runauthor {%
S. {\sc Shirai}, H. {\sc Tsuchiura}, Y. {\sc Asano}, Y. {\sc Tanaka}, 
J. {\sc Inoue}Y.{\sc Tanuma} and S.{\sc Kashiwaya}
}

\title
{%
Josephson effect in d-wave superconductor junctions in 
a lattice model }

\author{%
Shota {\sc Shirai}$^{1}$
Hiroki {\sc Tsuchiura}$^{2}$,
Yasuhiro {\sc Asano}$^{3}$,
Yukio {\sc Tanaka}$^{1,4}$,
Jun-ichiro {\sc Inoue}$^{1}$, 
Yasunari {\sc Tanuma}$^{5}$ and 
Satoshi {\sc Kashiwaya}$^{6}$
}

\inst
{
$^{1}$ Department of Applied Physics, Nagoya 
University, Nagoya, 464-8603, Japan
\\
$^{2}$ SISSA, Via Beirut, 2-4 34014, Trieste, Italy
\\
$^{3}$ Department of Applied Physics, Hokkaido University, 
Sapporo 060-8628, Japan
\\
$^{4}$ Crest, Japan Science and Technology Corporation (JST), 
Nagoya, 464-8063, Japan
\\
$^{5}$
Institute of Physics,
Kanagawa University,
Rokkakubashi, Yokohama, 221-8686, Japan
\\
$^{6}$ National Institute of Advanced Industrial Science and Technology, 
Tsukuba, 305-8568, Japan}

\recdate
{\today}

\abst{%
Josephson current between two $d$-wave superconductors is calculated 
by using a lattice model. Here we consider two types of junctions, 
$i.e.$, the parallel junction and the mirror-type junction. 
The maximum Josephson current 
$(J_{c})$ shows a wide variety of temperature ($T$) dependence 
depending on the misorientation angles and the types of junctions.
When the misorientation angles are not zero, 
the Josephson current shows the low-temperature anomaly
because of a zero energy state (ZES) at the interfaces. 
In the case of mirror-type junctions, $J_c$ has a non monotonic temperature
dependence.
These results are consistent with the previous results based on the quasiclassical theory. 
[Y.~Tanaka and S.~Kashiwaya: Phys. Rev. B \textbf{56} (1997) 892.] 
On the other hand, we find that the ZES disappears in several junctions
because of the Freidel oscillations of the wave function, which is peculiar to the 
lattice model. In such junctions, the temperature dependence of $J_{c}$ is close to 
the Ambegaokar-Baratoff relation. 
 }
\kword{
zero-energy states, $\pi$-junction
}
\begin{document}
\maketitle
\section{Introduction}
The Josephson effect is a supercurrent flow between superconductors, 
where the tunneling effect of Cooper pairs arises the electric 
current~\cite{Josephson}. 
This effect is quite distinct from other quasiparticle transport phenomena
in the sense that the macroscopic phase difference between the 
two superconductors plays an essential role. 
So far several expressions of the Josephson current have been derived
depending on the transport regimes of the region sandwiched by two
superconductors. Ambegaokar and Baratoff first derived a well known 
expression of the Josephson current in superconductor/insulator/superconductor (SIS) 
junctions in the tunneling limit (AB theory)~\cite{Ambegaokar}. 
Next Kulik and Omelyanchuk presented a expression available 
for superconductor/orifice/superconductor (SOS) junction \cite{Kulik1}. 
Then Josephson current in superconductor /normal metal /
superconductor (SNS) junctions was studied in several 
papers~\cite{Kulik2,Ishii,Bardeen,likharev}. 
After these works, Furusaki and Tsukada derived a general formula 
which covers the all junctions above on an equal footing 
(referred to as FT formula)~\cite{Furusaki}.
The applicability of these works, however, are limited to the conventional
$s$-wave superconductor junctions.  
The FT formula was extended to various directions such as
spin-singlet unconventional superconductors~\cite{review}
and spin-triplet superconductors~\cite{asano01-3,asano02-2,asano02-3,asano03-1}. 

The physics of $d$-wave superconductivity has been a hot topic 
in solid state physics since the discovery of high $T_{c}$ cuprates.
In $d$-wave superconductor junctions, 
the zero-energy state (ZES)~\cite{Hu} is formed at junction interfaces
because of an anomalous interference effect of a 
quasiparticle~\cite{matsumoto,nagato,ohhashi,sign1,sign2,sign3,Zhu1,Stefana,Wu,Dong,Kalen,Higashi}. 
Since the unconventional symmetry of the pair potentials is of the essence 
in the ZES~\cite{Phase,review},
the ZES is not expected in conventional $s$-wave superconductors. 
One of the striking effects of the ZES is the zero-bias conductance peak 
(ZBCP) in normal-metal/unconventional superconductor junctions
\cite{review,Tana1}.
In hybrid structures consisting of high-$T_{c}$ superconductors, 
a number of experiments observed the 
ZBCP~\cite{Kashi95,Kashi96,Alff,Wei,Wang,Iguchi,Kashi2000,Sawa1,Sawa2}.
The ZES affects various transport properties through junctions of 
unconventional superconductors.
~\cite{Zhu,Kashi2,Zutic,Yoshida,Hirai,Yoshida2002} 
The ZES also gives anomalous behaviors of the  
Josephson current in $d$-wave superconductor 
junctions~\cite{TK1,TK2,TK3,t61,Kusakabe,asano01-1,asano01-2}. 
Tanaka and Kashiwaya developed a general theory of Josephson current 
(TK theory) in spin-singlet unconventional junctions.
In the TK theory, following three important points are taken into account: 
1)an internal phase of the pair potential which induces $\pi$-junction 
\cite{Harlin,Barone,Sigrist,Tanaka1994}, 
2)the multiple Andreev reflection, 
and 
3)the formation of the ZES~\cite{TK1,TK2}.  
They have predicted that the current-phase relation 
in $d$-wave junctions is drastically changed from that in the usual 
Ambegaokar-Baratoff theory~\cite{TK1,TK2,riedel,samanta,barash}. 
Stimulated by their theory, there are several related 
works appear in recent years \cite{Stefana2,Barash3}. 
Experiments by Il'ichev $et.$ $al.$ showed that the current-phase relation 
of grain boundary YBCO junctions actually exhibit pronounced 
deviation from a simple sinusoidal current-phase relation~\cite{Ilichev1}. 
The experiments accomplished on mirror-type 45 degree junctions showed the 
presence of significant amount of $\sin(2\varphi)$ components, 
which is almost consistent with the TK theory. 
The anomalous enhancement of the Josephson current in low temperatures
and non monotonic temperature dependence of the Josephson current were 
also predicted by the TK theory. The latter has 
 been also observed in a experiment~\cite{Ilichev2}. 
Although the TK theory is qualitatively consistent with 
several experiments, there are still several remaining problems~\cite{Arie}. 
In the TK theory, the quasiclassical approximation 
\cite{eilen,larkin,zaitsev,shelankov,bruder} is employed 
on the derivation of the Josephson current. 
In high-$T_{c}$ cuprates, however,
the validity of the quasiclassical approximation may be questionable
since the coherence length (a few nm) is not much larger than
the Fermi wavelength.  
Thus we must check the validity of the TK theory
in a reasonable way. 
Actually a paper reported that the atomic scale roughness 
drastically influences the ZES~\cite{Tanuma1998}. 
In some cases, the Friedel oscillations of the wave function 
wash out the zero-energy peak (ZEP) in the local density of states 
near the interfaces~\cite{Tanuma1998}.
So far, however, such effects on the Josephson current have never been studied yet.
In order to address these issues, we have developed a theory of Josephson effect 
where the Bogoliubov-de Gennes equation is solved
numerically on a tight-binding lattice
~\cite{Hogan,Kawai,Tsuchiura,Shirai,asano01-1}. 
We calculate the current-phase relation and the temperature dependence of 
maximum Josephson current, $J_{c}$, for various misorientation angles 
in both the parallel and the mirror-type junctions. 
We have verified that the main conclusions in the TK theory hold, i.e.,
the large enhancement of $J_{c}$, the anomalous current-phase relation 
and the non-monotonous temperature dependence of $J_{c}$. 
In addition to this, we find the absence of the ZES in some junctions.
The Fermi wavelength characterizes the oscillations of the wave function.
The interference effect of a quasiparticle 
originating from such rapid oscillations
is responsible for the disappearance of the ZES.  
 In these junctions, the current-phase relation and temperature dependence of $J_{c}$ 
are almost explained by the AB theory. 
These results may serve as  a guide to fabricate high-$T_{c}$ 
junctions and predict a novel functionality 
originating from the formation of the ZES at the interface.~
\cite{Hilgen1,Tafuri,Smilde,Hilgen2,Imaizumi}  

The organization of this theory is as follows. 
In \S~2, the model and formulation is introduced. 
In \S~3, we show the numerical results of the parallel junctions. 
Corresponding results of the mirror-type junctions are introduced in \S~4. 
In \S~5, we summarize this paper.

\section{Model and Formulation}
Let us consider the extended-Hubbard Hamiltonian 
on two-dimensional tight-binding
model,
\begin{align}
H=& - \sum_{\boldsymbol{r},\boldsymbol{\rho},\sigma}
t_{\boldsymbol{r},\boldsymbol{r}+\boldsymbol{\rho}}
( c_{\boldsymbol{r},\sigma}^\dagger c_{\boldsymbol{r}+\boldsymbol{\rho},\sigma} ) 
-\mu \sum_{\boldsymbol{r},\sigma}c_{\boldsymbol{r},\sigma}^\dagger 
c_{\boldsymbol{r},\sigma}
-W/2\sum_{\boldsymbol{r},\boldsymbol{\rho},\sigma,\sigma'}
\left[n_{\boldsymbol{r},\sigma}n_{\boldsymbol{r}+\boldsymbol{\rho},\sigma'}\right],\label{exh}
\end{align}
where $\boldsymbol{r}=j\hat{\boldsymbol{x}}+m\hat{\boldsymbol{y}}$ 
labels a lattice site, $c_{\boldsymbol{r},\sigma}^\dagger$
($c_{\boldsymbol{r},\sigma}$) is the creation (anhilation) 
operator of an electron
at $\boldsymbol{r}$ with spin $\sigma$, 
$n_{\boldsymbol{r},\sigma}$ is the number operator. 
We assume the attractive interaction among the nearest neighbor sites (i.e., $W>0$).
In order to discuss Josephson effect in $d$-wave superconductivity, 
we apply the Hartree-Fock-Bogoliubov mean-field approximation.
The mean-field Hamiltonian reads,
\begin{align}
H_{MF}=& - \sum_{\boldsymbol{r},\boldsymbol{\rho},\sigma}
\left\{ t_{\boldsymbol{r},\boldsymbol{r}+\boldsymbol{\rho}} 
- \zeta_{\boldsymbol{r},\boldsymbol{r}+\boldsymbol{\rho}}\right\}
( c_{\boldsymbol{r},\sigma}^\dagger c_{\boldsymbol{r}+\boldsymbol{\rho},\sigma} ) 
-\tilde{\mu} \sum_{\boldsymbol{r},\sigma}c_{\boldsymbol{r},\sigma}^\dagger 
c_{\boldsymbol{r},\sigma}\nonumber\\
&+\sum_{\boldsymbol{r},\boldsymbol{\rho}}\left[
\Delta_{\boldsymbol{r},\boldsymbol{r}+\boldsymbol{\rho}}
c^\dagger_{\boldsymbol{r},\uparrow}c^\dagger_{\boldsymbol{r}+\boldsymbol{\rho},\downarrow}
+ h.c. \right]
+ E_0,\\
E_0 =& W/2 \sum_{\boldsymbol{r},\boldsymbol{\rho},\sigma,\sigma'} 
\langle n_{\boldsymbol{r},\sigma}\rangle 
\langle n_{\boldsymbol{r}+\boldsymbol{\rho},\sigma'}\rangle
-W/2 \sum_{\boldsymbol{r},\boldsymbol{\rho},\sigma} 
\langle c^\dagger_{\boldsymbol{r},\sigma}c_{\boldsymbol{r}+\boldsymbol{\rho},\sigma}\rangle 
\langle c^\dagger_{\boldsymbol{r}+\boldsymbol{\rho},\sigma}c_{\boldsymbol{r},\sigma}\rangle 
\nonumber\\
&+W \sum_{\boldsymbol{r},\boldsymbol{\rho}}
 \langle c^\dagger_{\boldsymbol{r},\uparrow}c^\dagger_{\boldsymbol{r}+\boldsymbol{\rho},\downarrow}
\rangle 
\langle c_{\boldsymbol{r}+\boldsymbol{\rho},\downarrow}c_{\boldsymbol{r},\uparrow}\rangle 
,\label{bcs}\\
\tilde{\mu}=& \mu + W \sum_{\boldsymbol{\rho},\sigma} \left\langle n_{\boldsymbol{\rho},\sigma}
\right\rangle, \\
\zeta_{\boldsymbol{r},\boldsymbol{r}+\boldsymbol{\rho}} =& W\left\langle 
c_{\boldsymbol{i},\sigma}^\dagger c_{\boldsymbol{i}+\boldsymbol{\rho},\sigma}
\right\rangle, 
\end{align}
In this paper, we take units of $\hbar=k_B=1$, where $k_B$ is the Boltzmann
constant.
The vectors $\boldsymbol{\rho}$ represents four vectors connecting 
nearest neighbor sites. The hopping integral  
between the nearest neighbor
sites is denoted by $t_{\boldsymbol{r},\boldsymbol{r}+\boldsymbol{\rho}}$. 
We define the pair potential 
\begin{equation}
\Delta_{\boldsymbol{r},\boldsymbol{r}+\boldsymbol{\rho}}
=-W \left\langle c_{\boldsymbol{r}+\boldsymbol{\rho},\downarrow} 
c_{\boldsymbol{r},\uparrow} \right\rangle.
\end{equation}

\begin{figure}
\begin{center}
\includegraphics[width=8.0cm]{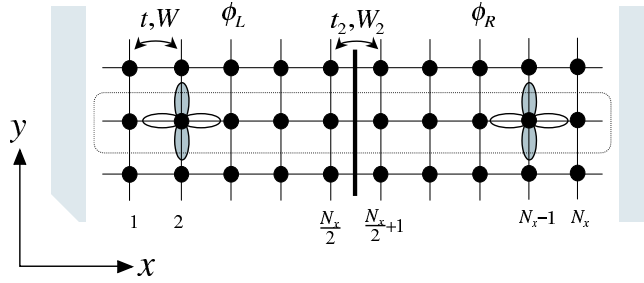}
\end{center}
\caption{ Schematic figures of the (100) parallel junctions. 
The pair potential is illustrated on the square lattice which represents 
the CuO$_2$ plane.
The solid line is the insulating barrier. The lattice sites surrounded 
by the dotted line corresponds to the unit cell. 
}
\label{fig:system}
\end{figure}

In the following, we explain the method to calculate the Josephson current
in a situation where the $a$ axes of two superconductors are perpendicular 
to the interface. Such junctions are referred to as the (100) parallel junction
in \S~2.
The number of the lattice sites in the $x$ direction is $N_x$ as shown in 
Fig.~\ref{fig:system}.
The periodic boundary condition is assumed in the $y$ direction.
The number of the lattice sites in the $y$ direction is $N_y N_c$, where
$N_y$ is the number the lattice sites included in a unit cell and $N_c$ 
is the number of the unit cells in the $y$ direction, respectively.
In the case of the (100) parallel junctions, $N_y$ =1.
The lattice sites included in the unit cell is surrounded by the dotted line 
in Fig.~\ref{fig:system}.
The hopping integral in superconductors 
is taken to be a constant $t$ and
that at the potential barrier is $t_2 e^{i\varphi/2}$, 
where $\varphi=\varphi_L-\varphi_R$ is the phase difference between 
the two superconductors.
We apply the Fourier transformation in the $y$ direction,
\begin{equation}
c_{j\hat{\boldsymbol{x}}+m\hat{\boldsymbol{y}},\sigma}=\frac{1}{N_c}
\sum_{k}c_{j,\sigma}(k)e^{ikm}.\label{ft1}
\end{equation}
The Hamiltonian in Eq.~(\ref{bcs}) results in
\begin{align}
 H_{MF}
=& \sum_{k}\sum_{j,j'=1}^{N_x}
\left[ c^\dagger_{j,\uparrow}(k), c_{j,\downarrow}(-k)\right]
\left[ \begin{array}{cc} h_0(j,j') & h_d(j,j') \\
                 h_d^\ast(j,j') & -h_0^\ast(j,j')
\end{array}
\right] 
\left[ \begin{array}{c} c_{j',\uparrow}(k) \\ c^\dagger_{j',\downarrow}(-k)
\end{array}\right] + E_0,\\
h_0(j,j') =& -(t+\zeta_1+\zeta_2) f(j,j')-\tilde{\mu}\delta_{j,j'}
- (\zeta_1 -\zeta_2) 2\cos k 
\nonumber\\
&-t_2 e^{i\varphi/2} \delta_{j,\frac{N_x}{2}+1} \delta_{j',\frac{N_x}{2}}
 -t_2 e^{-i\varphi/2} \delta_{j,\frac{N_x}{2}} \delta_{j',\frac{N_x}{2}+1},\\
h_d(j,j') =& \Delta f(j,j') -\Delta \delta_{j,j'} 2\cos k,\\
f(j,j') =& \delta_{j,j'+1}(1-\delta_{j',\frac{N_x}{2}})(1-\delta_{j',N_x})
+ \delta_{j,j'-1}(1-\delta_{j',\frac{N_x}{2}-1})(1-\delta_{j',1}),
\end{align}
By numerically solving the Bogoliubov-de Gennes (BdG) equation
\begin{align}
\sum_{j'=1}^{N_x}
\left[ \begin{array}{cc} h_0(j,j') & h_d(j,j') \\
                 h_d^\ast(j,j') & -h_0^\ast(j,j')
 \end{array}
\right] 
\left[ \begin{array}{c} u_{k,\lambda}(j') \\ v_{k,\lambda}(j') 
\end{array}\right] =E_{k,\lambda}
\left[ \begin{array}{c} u_{k,\lambda}(j) \\ v_{k,\lambda}(j) 
\end{array}\right],
\end{align}
the Hamiltonian in Eq.~(\ref{bcs}) can be diagonalized,
where $a_{k,\sigma,\lambda}^\dagger$ is the creation operator of a Bogoliubov
quasiparticle.
Throughout this paper, we assume the $d$-wave
symmetry in the pair potential and neglect the spatial dependence of the 
pair potential. 
Thus
\begin{equation}
\Delta_{\boldsymbol{r},\boldsymbol{r}+\boldsymbol{\rho}} = 
\left\{
\begin{array}{ccc}
\Delta &:& \boldsymbol{\rho} = \pm \hat{\boldsymbol{x}}\\
-\Delta&: &\boldsymbol{\rho} = \pm \hat{\boldsymbol{y}}
\end{array}
\right.,
\end{equation}
where the amplitude of $\Delta$ is determined by the gap equation
for the bulk superconductor in the $d$-wave symmetry~\cite{tsuchiura95}
\begin{align}
1=& \frac{W}{4} \frac{1}{N} \sum_{\boldsymbol{q}}
\frac{\gamma_{\boldsymbol{q}}^2}{E_{\boldsymbol{q}}}
\tanh \frac{E_{\boldsymbol{q}}}{2T},\\
\epsilon_{\boldsymbol{q}}=&-(t+\zeta_1)\eta_{\boldsymbol{q}} - \zeta_2\gamma_{\boldsymbol{q}}
-(\mu+8W n), \\
E_{\boldsymbol{q}} =&\sqrt{ \epsilon_{\boldsymbol{q}}^2 + \Delta^2
\gamma_{\boldsymbol{q}}^2},\\
\gamma_{\boldsymbol{q}} =& 2 ( \cos q_x - \cos q_y),\\
\eta_{\boldsymbol{q}} =& 2( \cos q_x + \cos q_y). 
\end{align}
In the same way, $\zeta_1$, $\zeta_2$ and $n$ are determined by the self-consistent equations,
\begin{align}
\zeta_1=& -\frac{W}{8} \frac{1}{N} \sum_{\boldsymbol{q}}\eta_{\boldsymbol{q}}
\frac{\epsilon_{\boldsymbol{q}}}{E_{\boldsymbol{q}}}
\tanh \frac{E_{\boldsymbol{q}}}{2T},\\
\zeta_2=& -\frac{W}{8} \frac{1}{N} \sum_{\boldsymbol{q}}\gamma_{\boldsymbol{q}}
\frac{\epsilon_{\boldsymbol{q}}}{E_{\boldsymbol{q}}}
\tanh \frac{E_{\boldsymbol{q}}}{2T},\\
n=& \frac{1}{N} \sum_{\boldsymbol{q}}
\left(1-\frac{\epsilon_{\boldsymbol{q}}}{E_{\boldsymbol{q}}}
\tanh \frac{E_{\boldsymbol{q}}}{2T}\right).
\end{align}
First, we detemine the magnitude of $\Delta$, $\zeta_1$, $\zeta_2$, and $\mu$ for $t=W$ in 
infinite $d$-wave superconductor so that 
$n=\langle \sum_{\sigma}n_{\boldsymbol{r},\sigma} \rangle =0.85$ 
is satisfied. Then the Josephson current is calculated by using these parameters.

The local density of states is defined by
\begin{equation}
N_j(E) = \frac{1}{N_c} \sum_{k,\lambda}\left\{ |u_{k,\lambda}(j)|^2 \delta( E_{k,\lambda}-E)
 +|v_{k,\lambda}(j)|^2 \delta( E_{k,\lambda}+E) \right\}.
\end{equation}
The bulk density of states is also given in this equation with $j$ being far from
both the interface and the edge of superconductors.
The free energy of the junction is calculated to be
\begin{align}
F(\varphi) =& -T \ln Z,\\
Z=& \textrm{Tr} \exp(-H_{MF}/T).
\end{align}
The Josephson current is determined by an equation
\begin{equation}
J(\varphi) = J = {2e} \frac{\partial F(\varphi)}{\partial \varphi}.
\end{equation}

The application of the method to other parallel and mirror-type
juntions is straightforward.

\section{Numerical Results in parallel junctions}

\begin{figure}
\begin{center}
\includegraphics[width=8.0cm]{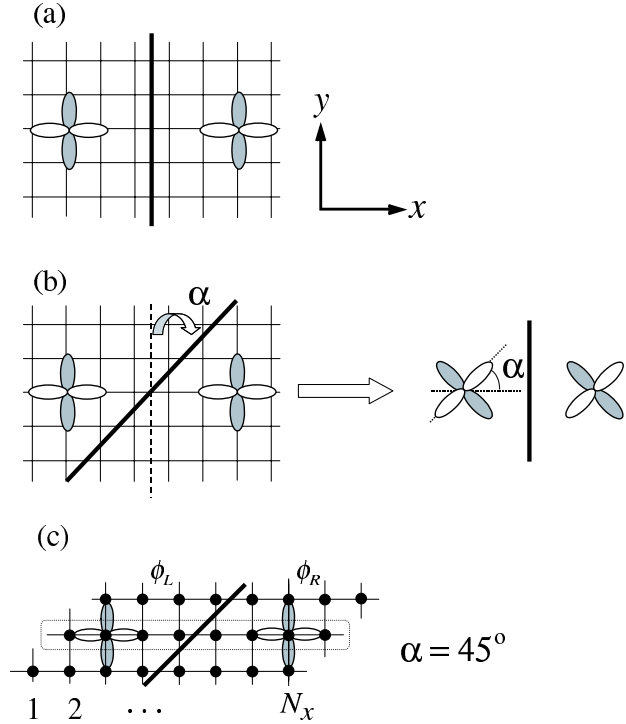}
\end{center}
\caption{ Schematic figures of the parallel junctions. The pair potential
is illustrated on the square lattice which represents the CuO$_2$ plane.
The solid line is the insulating barrier introduced on the
CuO$_2$ plane, where $\alpha$ is the orientation angle between the 
insulating layer and the $a$ axis of high-$T_c$ materials.
}
\label{fig:p1}
\end{figure}

The parallel junctions can be fabricated by introducing an insulating barrier
onto the CuO$_2$ plane as shown in
Fig.~\ref{fig:p1}. 
When the insulating barrier is parallel to the $y$ direction,
we obtain the (100) parallel junctions as shown in Fig.~\ref{fig:p1} (a).
The angle between the insulating layer and the $y$ direction is the orientation 
angle $\alpha$ as shown in Fig.~\ref{fig:p1} (b). 
Effects of insulating barrier is taken into account through 
the hopping integral across the barrier, $t_2=0.05 t$. 
When we calculate the free energy, the size of the unit cell in $x$ direction is 
taken to be $N_x=100$ for all junctions in this paper.
The size of the unit cell in the $y$ direction, $N_y$, depends on $\alpha$.
In the case of $\alpha =45^{\textrm{o}}$, for instance, $N_y = 1$ as shown in 
Fig.~\ref{fig:p1} (c), where the 
lattice sites surrounded by the dotted line correspond to the unit cell.
The number of unit cells in the $y$ direction is fixed at $N_c=200$ which
corresponds to the numer of $k$ in the summation of Eq.~(\ref{ft1}).
In the following, we show the calculated results in (100), (110), (120) and (130)
parallel junctions, where $\alpha= 0^\textrm{o}$, $45^{\textrm{o}}$, $26.5^{\textrm{o}}$ and 
$18.4^{\textrm{o}}$, respectively.  

\begin{figure}
\begin{center}
\includegraphics[width=10.0cm]{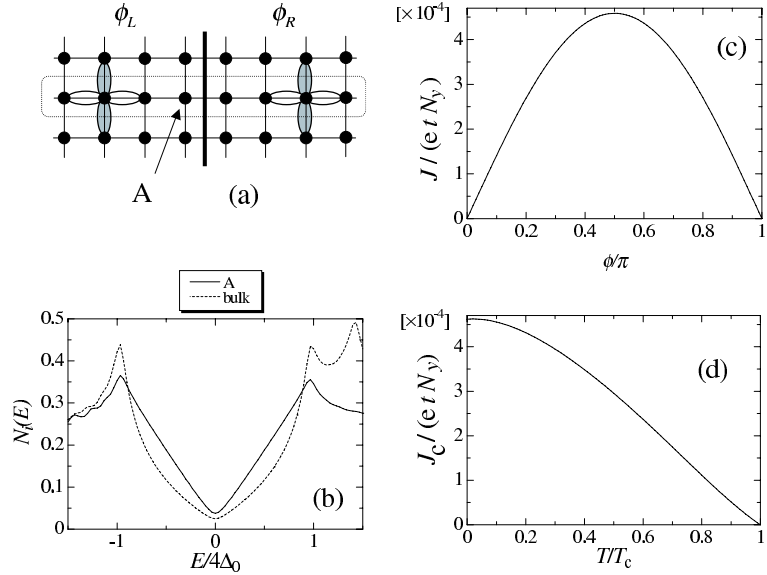}
\end{center}
\caption{ The (100) parallel junction is illustrated in (a).
The local density of states at a lattice site $A$ and the bulk density
of states are shown in (b).
The current-phase relation at $T=0$ and the temperature dependence of $J_c$ 
are shown in (c) and (d), respectively.
}
\label{fig:p2}
\end{figure}

In Fig.~\ref{fig:p2} (a), we illustrate the (100) parallel junction, where 
$\alpha=0^{\textrm{o}}$ and the unit cell used in the calculation is indicated by the 
dotted line.
In Fig.~\ref{fig:p2} (b), we show the local density of state 
at the lattice site $A$ and the bulk density of states, where 
$\varphi=\pi$ and $T=0$. The horizontal axis is the energy of a quasiparticle measured 
from the Fermi energy, where $\Delta_0$ is the amplitude of the pair
potential at the zero-temperature. There is no peak at the zero-energy in 
the local density of states. The results indicate no ZES at the interface,
which are consistent with the TK theory.
In Fig.~\ref{fig:p2} (c) and (d), we show the current-phase relation at $T=0$ 
and the maximum Josephson current as a function of temperatures, respectively.
The Josephson current is proportional to $\sin\varphi$ and takes its maximum at
$\varphi = 0.5\pi$ as shown in (c).
The temperature dependence of $J_c$ is described well by the Ambegaokar-Baratoff
relation as shown in (d).

\begin{figure}
\begin{center}
\includegraphics[width=10.0cm]{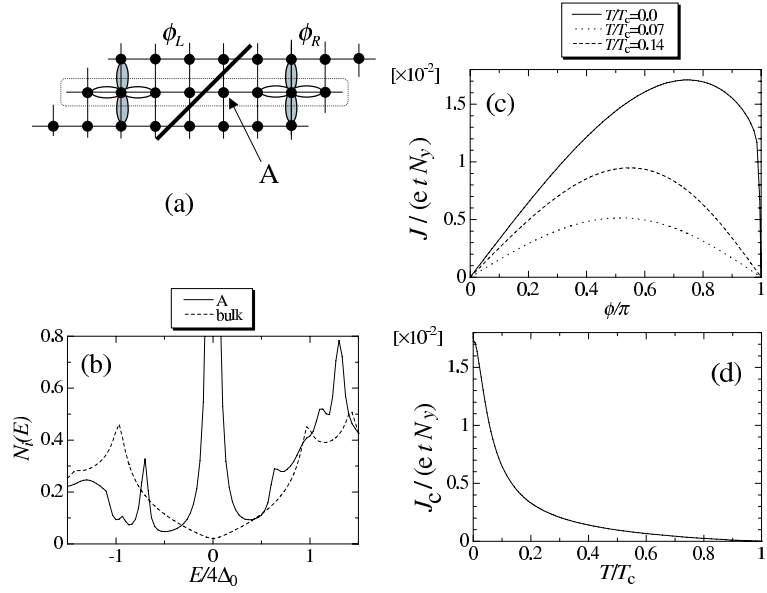}
\end{center}
\caption{ The (110) parallel junction is illustrated in (a).
The local density of states at a lattice site $A$ and the bulk density
of states are shown in (b).
The current-phase relation and the temperature dependence of $J_c$ 
are shown in (c) and (d), respectively.
}
\label{fig:p3}
\end{figure}

In Fig.~\ref{fig:p3}, we illustrate the (110) parallel junction in (a), where 
$\alpha=45^\textrm{o}$, $N_y=1$ and the unit cell used in the calculation is indicated by 
the dotted line.
In Fig.~\ref{fig:p3} (b), we show the local density of state 
at the lattice site $A$ and the bulk density of states, where 
$\varphi=\pi$ and $T=0$. There is a large zero-energy peak (ZEP) in 
the local density of states at $A$, whereas there is no ZEP in
the bulk density of states. The results indicate the presence of the ZES near the interface, 
which affects the Josephson current.
In Fig.~\ref{fig:p3} (c) and (d), we show the current-phase relation for several $T$
and $J_c$ as a function of temperatures, respectively.
In low temperatures, the Josephson current deviates from the sinusoidal function
of $\varphi$ because the resonant tunneling via the ZES enhances the 
transmission of Cooper pairs, which results in the multiple Andreev reflection. 
As a consequence, the current-phase relation becomes similar to that in SOS 
junctions~\cite{Kulik2}. The Josephson current takes its maximum at 
$\varphi=0.75\pi$ at $T=0$.
In Fig.~\ref{fig:p3} (d), $J_c$ increases rapidly with decreasing temperatures, which 
is called the low-temperature anomaly. The ZES is responsible for
the low-temperature anomaly in the Josephson current. 
These results are consistent with the TK theory.

\begin{figure}
\begin{center}
\includegraphics[width=10.0cm]{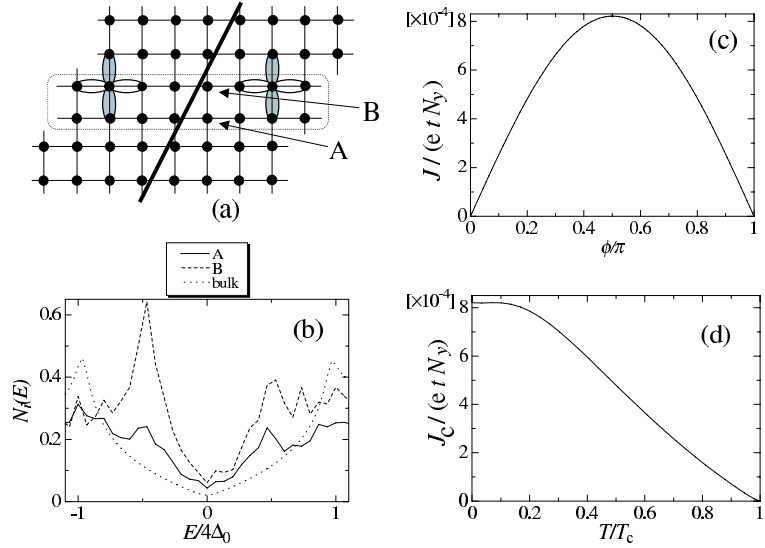}
\end{center}
\caption{ The (120) parallel junction is illustrated in (a).
The local density of states at $A$ and $B$, and the bulk density
of states are shown in (b).
The current-phase relation and the temperature dependence of $J_c$ 
are shown in (c) and (d), respectively.
}
\label{fig:p4}
\end{figure}

In Fig.~\ref{fig:p4}, we illustrate the (120) parallel junction in (a), where 
$\alpha=26.5^\textrm{o}$, $N_y=2$ 
and the unit cell used in the calculation is indicated by the dotted line.
There are two different lattice sites at the interface. 
From the site $A$, the hopping in $-x$ direction goes across the insulating barrier.
On the other hand, the hopping in $-x$ and $+y$ directions go beyond the insulator from 
the site $B$ as shwon in Fig.~\ref{fig:p4} (a).
In Fig.~\ref{fig:p4} (b), we show the local density of state 
at the lattice sites $A$ and $B$, where 
$\varphi=\pi$ and $T=0$. For comparison, we also show the bulk density of states.
According to the TK theory, a peak at the zero-energy is expected in the 
local density of states at both $A$ and $B$. The results in Fig.~\ref{fig:p4} (b), however,
do not show peak structures around the zero-energy, which contradicts 
to the TK theory. The disappearance of the ZES can be explained in terms of 
the Friedel oscillations of the wave function.
The period in the spatial oscillations of the wave function is 
 characterized by the Fermi wave length. 
The Fermi surface near the half-filling has almost the square
shape and the Fermi wave length corresponds to two lattice constants.
Thus the wave functions of the ZES at $A$ and $B$ cancelled with each
other. The absence of the ZES was also reported in the local density of states
at the (120) surface in the extended Hubbard model ~\cite{Tanuma1998} and 
$t-J$ model~\cite{TJ1,TJ2}. 
In Figs.~\ref{fig:p4} (c) and (d), we show the current-phase relation at $T=0$ 
and $J_c$ as a function of temperatures, respectively,
where the results are normalized by $N_y=2$.
The Josephson current shows the sinusoidal current-phase relation even at $T=0$.
This is because the resonant transmission of Cooper pairs is suppressed in the
 absence of the
ZES and the contributions of the multiple Andreev reflection are negligible.
In Fig.~\ref{fig:p4} (d), $J_c$ shows the saturation in low temperatures for $T<0.15T_c$, which 
are qualitatively same with those in the AB theory.
At the zero-temperature, $J_c$ in the (120) junction is as large as 1.8 times
of $J_c$ in the (100) junction. This is because three hopping pathes go over the barrier
in the unit cell of the (120) junctions, whereas the two superconductors are connected
by only one hopping path in the unit cell of the (100) junction.
In the (120) junctions, the ZES disappears because of the interference effect 
of a quasiparticle. This is a direct consequence of the electronic
structures in the lattice model near the half-filling.

\begin{figure}
\begin{center}
\includegraphics[width=10.0cm]{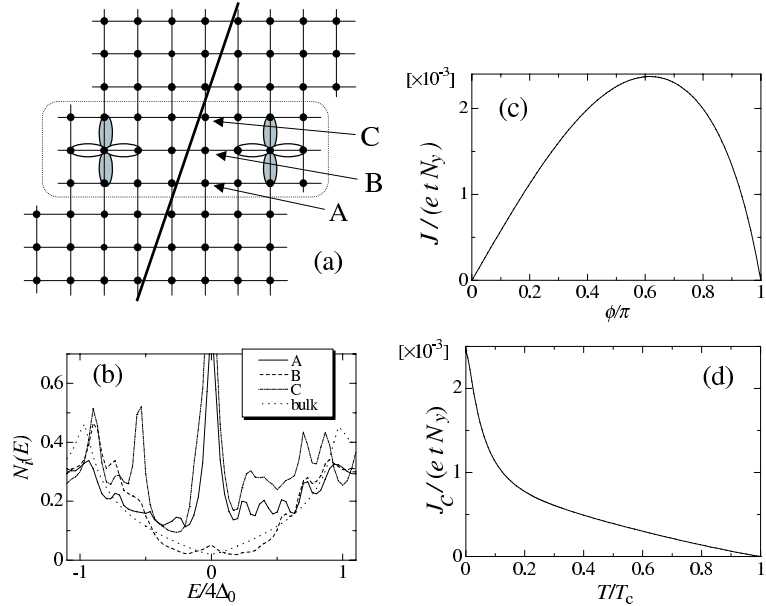}
\end{center}
\caption{ The (130) parallel junction is illustrated in (a).
The local density of states at lattice sites $A$, $B$ and $C$ are shown in (b).
The current-phase relation and the temperature dependence of $J_c$ 
are shown in (c) and (d), respectively.
}
\label{fig:p5}
\end{figure}

In Fig.~\ref{fig:p5}, we illustrate the (130) parallel junction in (a), where 
$\alpha=18.4^\textrm{o}$, $N_y=3$ and the unit cell used in the calculation 
is indicated by the dotted line.
There are three different lattice sites at the interface as indicated
by $A$, $B$ and $C$. 
In Fig.~\ref{fig:p5} (b), we show the bulk density of states and 
the local density of state 
at $A$, $B$ and $C$, where $\varphi=\pi$ and $T=0$. 
The local density of states at $A$ and $C$ shows a large peak at the zero-energy,
whereas there is no ZEP in the density of states at $B$. 
The absence of the ZES at $B$ can be also explained by the Friedel oscillations 
of the wave function. The waves propagating from $A$ and $C$ 
interfere destructively with that at $B$ in this case. 
As shown in the local density of states in the (120) and the (130) junctions,
the number of columns in the $y$ direction included in the unit cell, $N_y$, 
dominates the presence or the absence of the ZES.
When $N_y$ is odd integers, the ZES appears at the interface.
In the case of even integers, on the other hand, we find no ZES.
In Figs.~\ref{fig:p5} (c) and (d), we show the current-phase relation at $T=0$
and $J_c$ as a function of temperatures, respectively.
The Josephson current deviates from $\sin\varphi$ 
because of the resonant transmission of Cooper pairs 
through the ZES's at $A$ and $C$.
The degree of the deviation is rather small
when we compare the results in Fig.~\ref{fig:p5} (c) with those in Fig.~\ref{fig:p3}(c).
This is because the ZES is absent at $B$ and the degree of the resonance 
in the (130) junctions is weaker than that in the (110) junctions.
Actually, the Josephson current takes its maximum at $\varphi=0.61\pi$ 
in the (130) junctions, whereas $\varphi=0.75\pi$ characterizes the $J_c$ 
in the (110) junctions.
In Fig.~\ref{fig:p5} (d), $J_c$ shows the low-temperature anomaly as well as that in
the (110) junctions and increases rapidly with decreasing temperatures.

\section{Numerical Results in mirror-type junctions}
\begin{figure}
\begin{center}
\includegraphics[width=8.0cm]{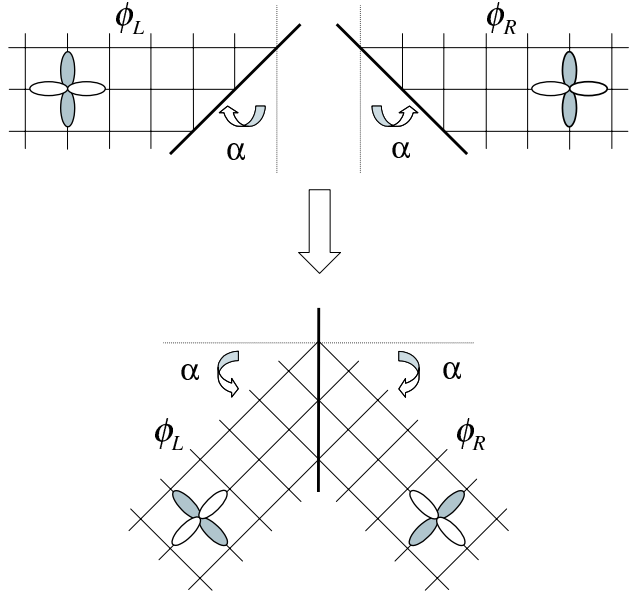}
\end{center}
\caption{ Schematic figures of the mirror type junctions. }
\label{fig:m1}
\end{figure}

To fabricate the mirror-type junctions, we first cut the two CuO$_2$ planes 
along the line oriented by $\alpha$ from the $a$ axis as shown in Fig.~\ref{fig:m1},
where the thick solid lines indicate the cutting lines.
Then we attach one cutting line to the other. 
The cutting line corresponds to the insulating barrier as shown in Fig.~\ref{fig:m1} (b).
In experiments, the mirror-type junction can be fabricated on 
bicrystal substrates, where the grain boundary formed between the 
superconductors on the different crystal axes insulates the 
two superconductors.
In the following, we show the calculated results in the (110), (120) and (130)
junctions as well as those in the  parallel junctions.
We note that the (100) mirror-type junction is essentially the same with 
the (100) parallel junction.

\begin{figure}
\begin{center}
\includegraphics[width=10.0cm]{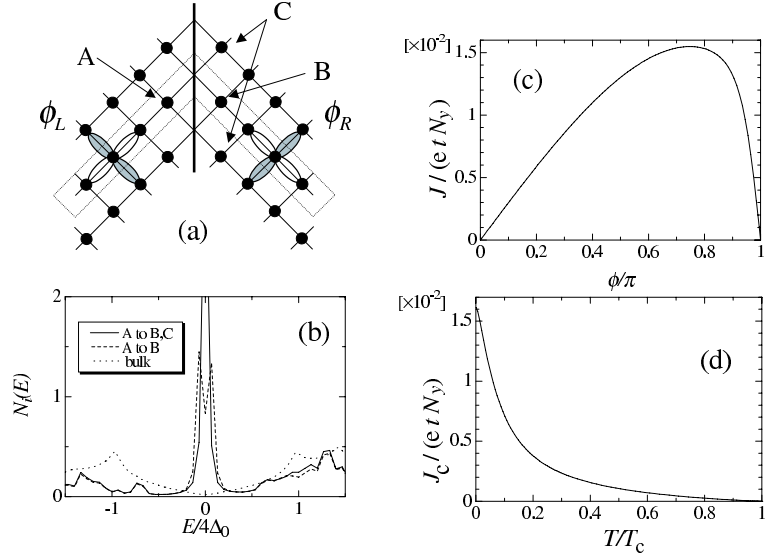}
\end{center}
\caption{ The (110) mirror type junction is illustrated in (a).
The local density of states at a lattice site $A$ and the bulk density
of states are shown in (b). We consider the hopping between $A$ and $C$ ($t_3$)
in the solid line and $t_3$ is set to be zero in the broken line.
The current-phase relation and the temperature dependence of $J_c$ 
are shown in (c) and (d), respectively.
}
\label{fig:m2}
\end{figure}

In Fig.~\ref{fig:m2}, we illustrate the (110) mirror-type junction in (a), 
where $\alpha=45^\textrm{o}$, $N_y=1$
and the unit cell used in the calculation is indicated by the dotted line.
In Fig.~\ref{fig:m2} (b), we show the local density of state 
at the lattice site $A$ and the bulk density of states, where 
$\varphi=\pi$ and $T=0$.
In addition to the hopping between $A$ and $B$, ($t_2$), we also consider
the hopping between $A$ and $C$, ($t_3=1/\sqrt{2}t_2$). The local 
density of states shows a large ZEP as show in the solid line.
These results are qualitatively the same 
with those in the (110) parallel junctions.
For comparison, we show the density of states at $A$
in the absence of the hopping between $A$ and $C$, (i.e., $t_3$=0).
In this case, the ZEP splits into two peaks as shown with the broken line.
It is known that the disordered potential near the interface splits the 
ZBCP~\cite{asano03-2}. 
Since the absence of $t_3$ is considered to be an imperfection
at the interface, the origin of the splitting in Fig~\ref{fig:m2} (b) 
can be explained in the same way with that found in the disordered 
junctions~\cite{asano03-2}. 
In Fig.~\ref{fig:m2} (c) and (d), we show the current-phase relation at $T=0$
and the maximum Josephson current as a function of temperatures, respectively.
At $T=0$, the Josephson current deviates from $\sin\varphi$ 
because the resonant tunneling via the ZES enhances the 
transmission of Cooper pairs. The Josephson current at $T=0$ takes its maximum
at $\varphi=0.75\pi$ as well as that in the (110) parallel junction.
In Fig.~\ref{fig:m2} (d), $J_c$ shows the low-temperature anomaly 
because of the ZES. We note that the Josephson current in the (110) mirror-type junction 
flows opposite direction to that in the (110) parallel junction. 
These results are consistent with the TK theory.

\begin{figure}
\begin{center}
\includegraphics[width=10.0cm]{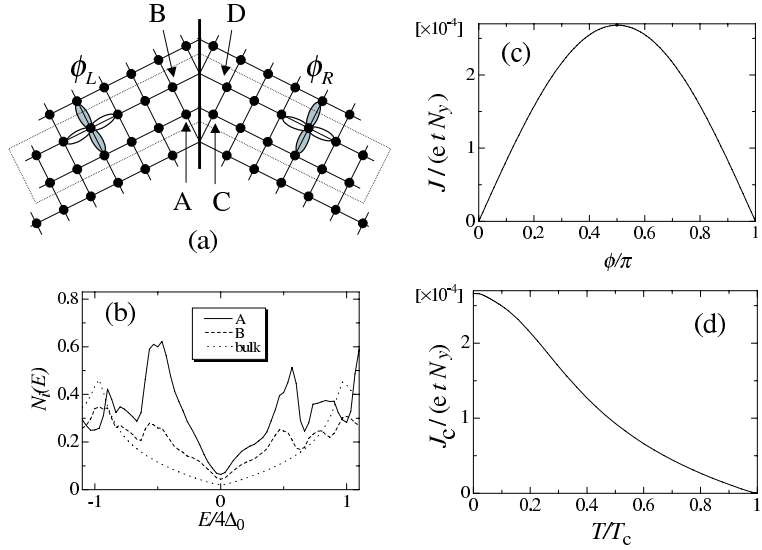}
\end{center}
\caption{ The (120) mirror type junction is illustrated in (a).
The local density of states at $A$ and $B$, and the bulk density
of states are shown in (b).
The current-phase relation and the temperature dependence of $J_c$ 
are shown in (c) and (d), respectively.
}
\label{fig:m3}
\end{figure}

In Fig.~\ref{fig:m3}, we illustrate the (120) mirror type junction in (a), 
where $\alpha=26.5^\textrm{o}$, $N_y=2$ 
and the unit cell used in the calculation is indicated by the dotted line.
There are four different lattice sites near the interface as indicated by $A$, 
$B$, $C$ and $D$. 
In addition to the hopping between $A$ and $C$ ($t_2$), we also consider
the hopping between $B$ and $D$, where we assume $t_3=t_2/2$. 
In Fig.~\ref{fig:m3} (b), we show the local density of state 
at $A$ and $B$, where 
$\varphi=\pi$ and $T=0$. For comparison, we also show the bulk density of states.
The results show the absence of the ZEP, which contradicts to the TK theory.
The disappearance of the ZES in this case can be also explained in the same 
way as that found in the (120) parallel junctions.
Since the Fermi wavelength corresponds to two lattice constants,
the wave functions of the ZES at $A$ and $B$ cancelled with each
other. In Figs.~\ref{fig:m3} (c) and (d), we show the current-phase relation at $T=0$ 
and the maximum Josephson current as a function of temperatures, respectively.
The Josephson current shows the sinusoidal current-phase relation even at $T=0$
because the resonant transmission of Cooper pairs is suppressed in the absence of the
ZES. In Fig.~\ref{fig:m3} (d), $J_c$ does not show the low-temperature anomaly.
In (120) mirror type junctions, the ZES also disappears because of the interference 
effect of a quasiparticle, which is one of characteristic features in the lattice model.

\begin{figure}
\begin{center}
\includegraphics[width=10.0cm]{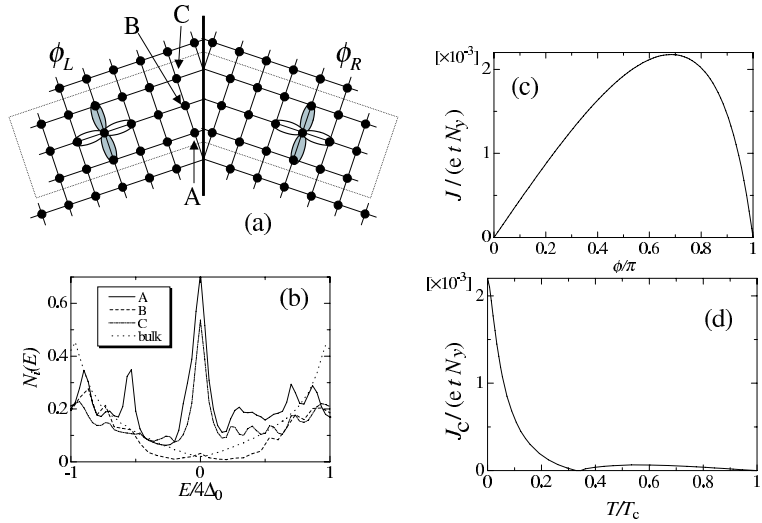}
\end{center}
\caption{ The (130) mirror type junction is illustrated in (a).
The local density of states at lattice sites $A$, $B$ and $C$ are shown in (b).
The current-phase relation and the temperature dependence of $J_c$ 
are shown in (c) and (d), respectively.
}
\label{fig:m4}
\end{figure}

In Fig.~\ref{fig:m4}, we illustrate the (130) mirror type junction in (a), 
where $\alpha=18.4^\textrm{o}$, $N_y=3$ 
and the unit cell used in the calculation is indicated by the dotted line.
There are three different lattice sites at the interface as indicated
by $A$, $B$ and $C$. In addition to the hopping in the $+x$ direction from 
$A$ ($t_2$),
we consider the hopping from $B$ ($t_3=t_2/2$) and from $C$ ($t_4=t_2/3$)
across the insulating barrier. 
In Fig.~\ref{fig:m4} (b), we show the the bulk density of states and 
the local density of state 
at $A$, $B$ and $C$, where $\varphi=\pi$ and $T=0$. 
The local density of states at $A$ and $C$ show a large ZEP,
whereas there is no ZEP in the density of states at $B$. 
The absence of the ZEP at $B$ is also 
explained by the Friedel oscillations of the wave functions.
In Figs.~\ref{fig:m4} (c) and (d), we show the current-phase relation at $T=0$
and $J_c$ as a function of temperatures, respectively.
The Josephson current deviates the sinusoidal function
of $\varphi$ because of the resonant tunneling of Cooper pairs via the ZES.
In Fig.~\ref{fig:m4} (d), $J_c$ vanishes at $T=0.33T_c$. 
Then $J_c$ rapidly increases with the decrease 
of temperatures. The minimum point of $F(\varphi)$ is changed from $\varphi=0$ 
to $\varphi=\pi$, when temperatures decreases across $0.33T_c$. Thus the 
junction becomes the 0-junction for $T>0.33T_c$ and the $\pi$-junction for $T<0.33T_c$.
At $T=0.33T_c$, the supercurrent flow changes its direction, which is a characteristic
behavior of the Josephson current in the mirror-type junctions.

\section{Conclusions}
In this paper, the Josephson current in $d$-wave superconductor / 
$d$-wave superconductor junctions is calculated based on 
a lattice model.  Here we consider two types of junctions, 
($i.e.$, the parallel junction and the mirror-type junction) with  
several misorientation angles. 
In both types of junctions, $J_{c}$ shows the low-temperature 
anomaly in the presence of the ZES at the junction interface.
In such situation, the current-phase relation deviates 
significantly from a sinusoidal function of $\varphi$. 
We find the ZES in the (110) and (130) junctions. 
In the (100) and (120) junctions, on the other hand, 
we find no ZES near the interface.
The theory by Tanaka and Kashiwaya (TK theory) predicted 
the presence of the ZES in the junctions other than the (100)
junctions. Thus the absence 
of the ZES in the (100) junction is consistent with the TK theory.
The calculated results in the 
(120) junction, however, contradict to the TK theory.
In the TK theory, the quasiclassical approximation is employed 
to derive the Josephson current formula. The approximation
is justified when the coherence length is much larger than the Fermi
wave length. The electronic structures in high-$T_c$ superconductors
may be described by those in the two-dimensional tight-binding
model near the half-filling. The coherence length is 
considered to be comparable to the Fermi wavelength. 
Indeed, the wave function of a quasiparticle 
at the zero-energy interferes destructively near the interface of 
the (120) junctions, which leads to the absence of the ZES.
This interference effect is peculiar to the tight-binding model
near the half-filling. 
Since there is no ZES at the interface, 
the current-phase relation becomes the sinusoidal function of $\varphi$ and 
the temperature dependence of the maximum Josephson current is
close to the results in the Ambegaokar-Baratoff theory. 
However, when the electron density per site 
deviates far away from the hal-filling, 
such destructive interference effect does not happen and ZES 
recovers \cite{Tanuma1998}. 
In this case, the Josephson current is expected 
to be consistent with TK theory. 
The characteristic behavior of the Josephson current in two types
of junctions is qualitatively different from each other.
In the (110) junctions, the direction of the supercurrent flow
in the parallel junction is opposite to that in the mirror-type 
junction.
Furthermore, we have found non-monotonic temperature 
dependence of $J_{c}$ in the (130) mirror-type junction. 
These results are consistent with the TK theory. 
In this study, we have confirmed that main conclusions  
of the TK theory: 
i) the enhancement of $J_{c}$ at low temperatures and 
ii)non monotonous temperature dependence of $J_{c}$,
are valid even if we consider much more realistic electronic
structures in high-$T_c$ materials. 

There are several remaining problems. 
In the present study, flat interfaces are assumed for the simplicity. 
Since random potentials near the interface suppress  
the ZES 
\cite{barash2,golubov,poenicke,yamada,luck,Tanaka2002,asano01-1,asano01-2,Itoh1,Itoh2,asano02-1,asano03-2,asano02-4}, 
it may be important to clarify effects of the 
atomic scale roughness on the Josephson current. 

In the present paper, the spatial depletion of the 
pair potential is not taken into account for simplicity. 
To our knowledge, this treatment would not seriously 
modify the conclusion of this paper 
unless subdominant components of the pair potential
do not break the time reversal symmetry~\cite{review}. 
When subdominant $s$ or $d_{xy}$ component 
breaks the time reversal symmetry near the interface, however,
the temperature dependence of $J_{c}$ would be 
seriously changed  by such broken time reversal symmetry 
state (BTRSS)~\cite{TK3}. 
Although there are several works about the BTRSS 
\cite{fogelstrom,YT021,YT022,YT023,Kashiwaya95,TJ1,TJ2,Tanuma2001,Zhu2,Amin,matsumoto2,laughlin,tanaka3,tanaka4,covington}, 
it has not been established yet if such state is really realized at the 
interface~\cite{biswas,dagan,sharoni,Ekin,Aubin,Neils}. 
However, from the theoretical view point, the extension of the present theory 
in this direction is a challenging future problem. 

In the present paper, we only focus on the dc Josephson effect 
at the zero bias-voltage across the junctions. 
It is also interesting to study the quasiparticle current and 
the ac Josephson effect~\cite{Hurd,Lofwander,Yontai}
in the present approach.

\acknowledgements 

This work was partially supported by the Core Research for
Evolutional Science and Technology (CREST) of the Japan Science and
Technology Corporation (JST). 
The computational aspect of this work has been performed at the 
facilities of the Supercomputer Center, Institute of Solid State 
Physics, University of Tokyo and the Computer Center. 
J.I. acknowledges support by the NEDO 
international Joint Research project 
"Nano-scale Magnetoelectronics". 
%
%

\end{document}